\documentclass{PoS}
\pdfoutput=1
\usepackage{bm}

\newcommand{\dvec}{\bm{d}}
\newcommand{\Pvec}{\bm{P}}
\newcommand{\beq}{\begin{equation}}
\newcommand{\eeq}{\end{equation}}
\newcommand{\beqy}{\begin{eqnarray}}
\newcommand{\eeqy}{\end{eqnarray}}
\newcommand{\svec}{\bm{s}}
\newcommand{\nn}{\nonumber}
\newcommand{\qcm}{\bm{q}_{\rm cm}}
\newcommand{\qq}{\qcm^{2}}
\newcommand{\Ecm}{E_{\rm cm}}
\newcommand{\nvec}{\bm{n}}
\newcommand{\wvec}{\bm{w}}
\newcommand{\zvec}{\bm{z}}
\newcommand{\xvec}{\bm{x}}

\title{Pion-pion scattering phase shifts with the stochastic LapH method}

\ShortTitle{Pion-pion scattering}

\author{\speaker{Brendan Fahy}$^{a,b}$, John Bulava$^c$, 
        Ben H\"{o}rz$^c$, Keisuke~J.~Juge$^{d}$, 
        Colin Morningstar$^a$, and Chik Him~Wong$^e$\\
\llap{$^a$} Dept.~of Physics, Carnegie Mellon University, 
        Pittsburgh, PA 15213, USA\\
\llap{$^b$} High Energy Accelerator Research Organization (KEK), 
        Ibaraki 305-0801, Japan\\
\llap{$^c$} School of Mathematics, 
            Trinity College, Dublin 2, Ireland\\
\llap{$^d$} Dept.~of Physics, University of the Pacific, 
        Stockton, CA 95211, USA\\
\llap{$^e$} Department of Physics, University of Wuppertal, 
            Gaussstrasse 20, D-42119, Germany}

\abstract{Progress in calculating scattering phase shifts on $N_f=2+1$ anisotropic 
clover Wilson lattices is described. The stochastic LapH method facilitates 
computations in large volumes and for light pion masses. Results for pion masses 
down to 240 MeV, keeping $m_\pi L > 4$, are presented.}

\FullConference{The 32nd International Symposium on Lattice Field Theory,\\
		23-28 June, 2014\\
		Columbia University New York, NY}

\begin{document}

\section{Introduction}

Monte Carlo calculations in lattice QCD are necessarily carried out
in finite volume.  However, most of the excited hadrons we seek to study
are unstable resonances.  In finite volume with particular boundary
conditions, the eigenvalues of the Hamiltonian are discrete since
only certain momenta are allowed in order to satisfy the boundary
conditions.  Diagonalization of the Hamiltonian leads to a knowledge
of the discrete stationary states.  In infinite volume, a continuum
of momenta are available and unstable excited hadrons decay to multi-hadron
asymptotic states.  In finite volume, there are no decays; instead,
there is only quantum mechanical mixing between Fock states.
Fortunately, it is possible to study the excited resonances
using finite-volume calculations. 

The idea that finite-volume energies can be related to 
infinite-volume scattering processes is actually rather old, dating back to 
Refs.~\cite{dewitt1956a,DeWitt:1956b} in the mid-1950s.  Details on how
to utilize such relationships in lattice QCD were first spelled out in  
Refs.~\cite{luscher1991,rummukainen1995}.  These calculations
were later revisited using an entirely field theoretic approach
in Ref.~\cite{sharpe2005}, and subsequent works have generalized their results
to treat multi-channels with different particle masses and 
nonzero spins.

In the work described in this talk, we use a variety of two-pion energies 
in finite volume with different total momenta to calculate the $P$-wave 
scattering phase shifts in the $I=1$ channel, and extract the mass and width of
the $\rho$ resonance.  Our preliminary results are obtained on a
$32^3\times 256$ anisotropic lattice with quark masses tuned to yield
a pion mass around 240~MeV.  All needed Wick contractions are efficiently 
evaluated using a stochastic method\cite{StochasticLaph} of treating the 
low-lying modes of quark propagation that exploits Laplacian Heaviside 
quark-field smearing.

\section{Scattering phase shifts from finite-volume energies}

For a given total momentum $\Pvec=(2\pi/L)\dvec$ in a spatial $L^3$ volume
with periodic boundary conditions, where $\dvec$ is a vector of integers, 
we determine the total energy $E$ in the lab frame for a particular 
two-particle interacting state in our lattice QCD simulations.  If the 
masses of the two particles are $m_1$ and $m_2$, we then boost to the 
center-of-mass frame and define the following quantities:
\begin{eqnarray}
   \Ecm &=& \sqrt{E^2-\Pvec^2},\qquad
   \gamma = \frac{E}{\Ecm},\qquad
   \qq = \frac{1}{4} \Ecm^2
   - \frac{1}{2}(m_1^2+m_2^2) + \frac{(m_1^2-m_2^2)^2}{4\Ecm^2},\\
   u^2&=& \frac{L^2\qq}{(2\pi)^2},\qquad
 \svec = \left(1+\frac{(m_1^2-m_2^2)}{\Ecm^2}\right)\dvec.
\end{eqnarray}
The relationship between the finite-volume
two-particle energy $E$ and the infinite-volume scattering amplitudes
(and phase shifts) is encoded in the matrix equation:
\beq
   \det[1+F^{(\svec,\gamma,u)}(S-1)]=0,
\eeq
where $S$ is the usual $S$-matrix whose elements can be written in
terms of the scattering phase shifts, and the $F$ matrix is given in
the $JLS$ basis states by
\beq
F^{(\svec,\gamma,u)}_{J'm_{J'}L'S'a';\ Jm_JLSa} =
\frac{\rho_a}{2} \delta_{a'a}\delta_{S'S}\left\{\delta_{J'J}\delta_{m_{J'}m_J}
\delta_{L'L}+ W_{L'm_{L'};\ Lm_L}^{(\svec,\gamma,u)}
\langle J'm_{J'}\vert L'm_{L'},Sm_{S}\rangle
\langle Lm_L,Sm_S\vert Jm_J\rangle
\right\},
\label{eq:Fdef}
\eeq
where $J,J'$ refer to total angular momentum, $L,L'$ are total orbital angular
momenta, $S,S'$ refer to total intrinsic spin in the above equation, $a,a'$ label
channels, $\rho_a=1$ for distinguishable particles and $\rho_a=\frac{1}{2}$
for identical particles, and
\beq
W^{(\svec,\gamma,u)}_{L'm_{L'};\ Lm_L}
=\frac{2i}{\pi\gamma u^{l+1}}{\cal Z}_{lm}(
\svec,\gamma,u^2) \int\!d^2\Omega
\ Y^\ast_{L'm_{L'}}(\Omega) Y^\ast_{lm}(\Omega) Y_{Lm_L}(\Omega).
\label{eq:Wdef}
\eeq
Notice that $F^{(\svec,\gamma,u)}$ is diagonal in channel space, but mixes different
total angular momentum sectors, whereas $S$ is diagonal in angular
momentum, but has off-diagonal elements in channel space.  Also,
the matrix elements of $F^{(\svec,\gamma,u)}$ depend on the total momentum $\Pvec$
through $\svec$, whereas the matrix elements of $S$ do not.
The Rummukainen-Gottlieb-L\"uscher (RGL) shifted zeta functions are evaluated
using
\beqy
   {\cal Z}_{lm}(\svec,\gamma,u^2)&=&\sum_{\nvec\in \mathbb{Z}^3}
  \frac{{\cal Y}_{lm}(\zvec)}{(\zvec^2-u^2)}e^{-\Lambda(\zvec^2-u^2)}
 +\delta_{l0}\gamma\pi e^{\Lambda u^2}\left( 2u D(u\sqrt{\Lambda})
-\Lambda^{-1/2}\right)
\nn\\
 &+&\frac{i^l\gamma}{\Lambda^{l+1/2}} \int_0^1\!\!dt 
\left(\frac{\pi}{t}\right)^{l+3/2}\! e^{\Lambda t u^2}
\sum_{\nvec\in \mathbb{Z}^3\atop \nvec\neq 0}
e^{\pi i \nvec\cdot\svec}{\cal Y}_{lm}(\wvec)
\  e^{-\pi^2\wvec^2/(t\Lambda)},
\label{eq:zaccfinal}
\eeqy
where $\zvec= \nvec -\gamma^{-1} \bigl[\textstyle\frac{1}{2}
+(\gamma-1)s^{-2}\nvec\cdot\svec \bigl]\svec$ and
$\wvec=\nvec - (1  - \gamma) s^{-2}
 \svec\cdot\nvec\svec$, the spherical harmonic polynomials are given by
${\cal Y}_{lm}(\xvec)=\vert \xvec\vert^l\ Y_{lm}(\widehat{\xvec})$,
and $D(x)$ is the Dawson function, defined by
\beq
   D(x)=e^{-x^2}\int_0^x\!dt\ e^{t^2}.
\eeq
We choose $\Lambda\approx 1$, although the final answer is independent
of this choice.  Choosing $\Lambda$ near unity allows sufficient
convergence speed of the summations.  Gauss-Legendre quadrature is used to
perform the integral, and the Dawson function is evaluated using
a Rybicki approximation.

\begin{table}[t]
\begin{center}
\begin{tabular}{|ccc|} \hline
 $\dvec$ & $\Lambda$  & $\cot\delta_1$ \\ \hline\hline
(0,0,0)  & $T_{1u}^+$ & ${\rm Re}\ w_{0,0}$\\
(0,0,1)  & $A_1^+$    & ${\rm Re}\ w_{0,0}+\frac{2}{\sqrt{5}}{\rm Re}\ w_{2,0}$\\
         & $E^+$      & ${\rm Re}\ w_{0,0}-\frac{1}{\sqrt{5}}{\rm Re}\ w_{2,0}$\\
(0,1,1)  & $A_1^+$    & 
 $ {\rm Re}\ w_{0,0}+  \frac{1}{2\sqrt{5}} {\rm Re}\ w_{2,0}
                  -  \sqrt{\frac{6}{5}}  {\rm Im}\ w_{2,1}  - \sqrt{\frac{3}{10}} {\rm Re}\ w_{2,2},$\\
        & $B_1^+$    & 
 $ {\rm Re}\ w_{0,0}-\frac{1}{\sqrt{5}}{\rm Re}\ w_{2,0}
         + \sqrt{\frac{6}{5}} {\rm Re}\ w_{2,2} ,$\\
         & $B_2^+$    & 
 ${\rm Re}\ w_{0,0}+ \frac{1}{2\sqrt{5}}{\rm Re}\ w_{2,0}
      +\sqrt{\frac{6}{5}} {\rm Im} w_{2,1}-\sqrt{\frac{3}{10}}  {\rm Re}\ w_{2,2}$\\
(1,1,1)  & $A_1^+$    & ${\rm Re}\ w_{0,0}
                  + 2 \sqrt{\frac{6}{5}}  {\rm Im}\ w_{2,2}$\\
         & $E^+$      & ${\rm Re}\ w_{0,0} -\sqrt{\frac{6}{5}} {\rm Im}\ w_{2,2}$\\
\hline
\end{tabular}
\end{center}
\caption{Expressions for the $P$-wave phase shifts $\delta_1(\Ecm)$
relevant for $I=1$ $\pi\pi$ scattering for various $\dvec$ and irreps
$\Lambda$. The quantities $w_{lm}$ are defined 
in Eq.~(\protect\ref{eq:wdef}). The irrep labels are discussed
in Ref.~\protect\cite{ExtendedHadrons}.
\label{tab:cotdelta}}
\end{table}

The scattering processes we study conserve both total angular momentum $J$
and the projection of total angular momentum, say $M_J$.  Given orthonormal
states, then the unitarity of the $S$-matrix tells us that
\beq 
  \langle J'm_{J'}'L^\prime S^\prime a'\vert\ S
\ \vert Jm_JLS a\rangle = \delta_{J'J}\delta_{m_{J'}m_J}\ s^{(J)}_{L'S'a',\ LSa}(E),
\eeq
where $a',a$ denote other defining quantum numbers, such as channel, and
$s^{(J)}$ is a unitary matrix that is independent of $m_J$ due to rotational
invariance.  If the two
particles have zero spin $s_1=s_2=0$ and there is only one channel, then
\beq
   s^{(J)}=s^{(L)}=e^{2i\delta_L(E)},
\eeq
where $\delta_L(E)$ are known as the \textit{scattering phase shifts}.
The factor of 2 is conventional to agree with a certain definition when
scattering from a central potential.

For single-channel $\pi\pi$ scattering, $s_1=s_2=0$, so $S=0$ and $J=L$, 
in which case Eq.~(\ref{eq:Fdef}) simplifies to
\beq
  F^{(\svec,\gamma,u)}_{L'm_{L'};\ Lm_L} =
\frac{1}{2} \left(\delta_{L'L}\delta_{m_{L'}m_L}
+W_{L'm_{L'};\ Lm_L}\right),
\eeq
using $\rho_a=1$ for distinguishable pions. In the case of $P$-wave scattering
of pions, we assume $\delta_L=0$ for all $L$ except $L=1$. Hence, the matrix 
elements of $S-1$ are all zero, except for diagonal entries with $L=1$.  This 
means the matrix $F(S-1)$ has non-zero entries only for columns with $L=1$, so 
we only need to consider the $3\times 3$ block involving $L=1$.  In all cases,
we can reduce the $3\times 3$ matrix to diagonal form and obtain expressions 
for $\cot\delta_1$ for various $\dvec$ and 
irreps $\Lambda$, which are summarized in Table~\ref{tab:cotdelta}, defining
\beq
     w_{lm} = \frac{{\cal Z}_{lm}(\svec,\gamma,u^2)}{\gamma \pi^{3/2} u^{l+1}}.
\label{eq:wdef}
\eeq

\begin{figure}[t]
  \begin{center}
  \includegraphics[width=2.9in]{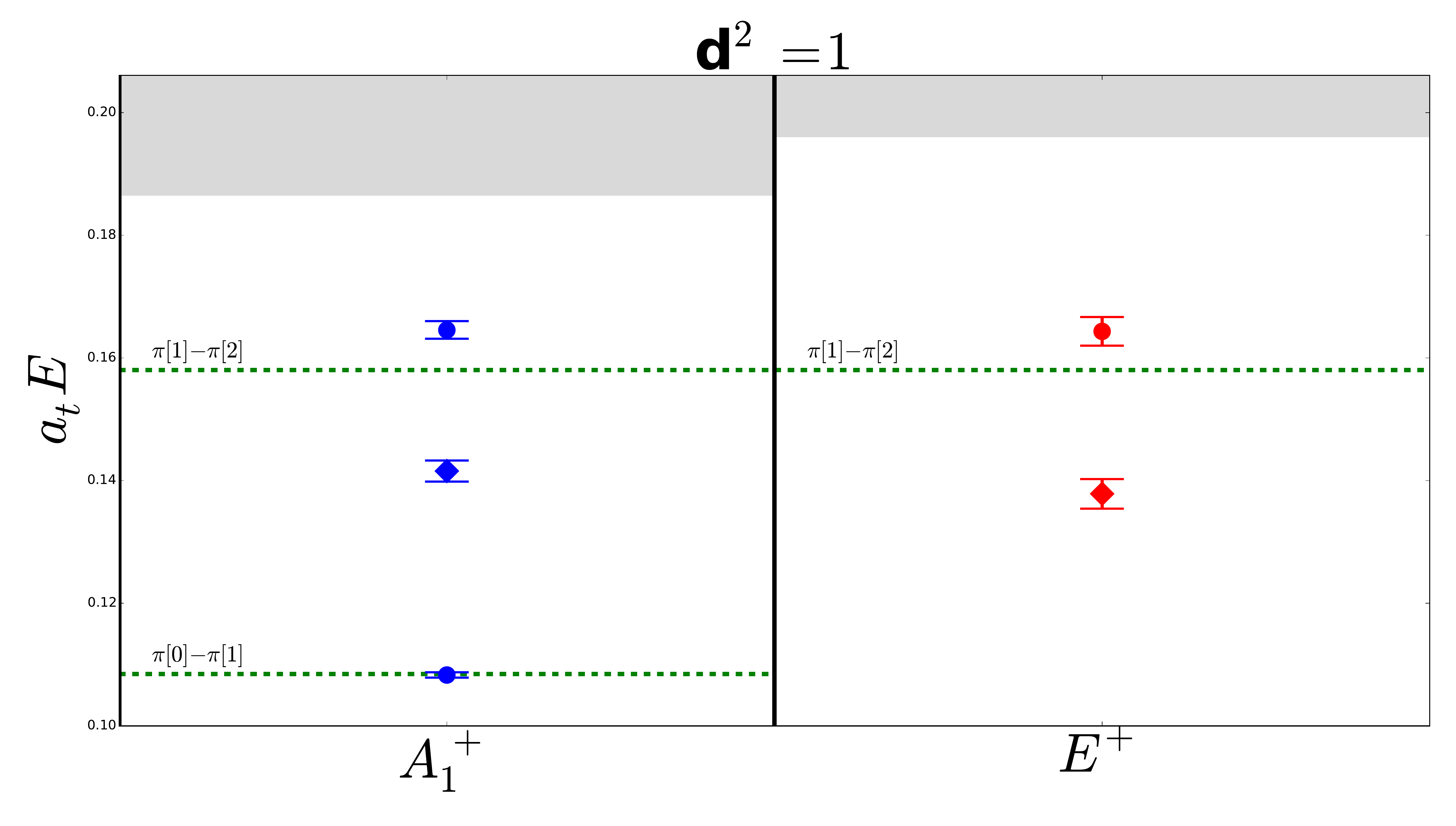}
  \includegraphics[width=2.9in]{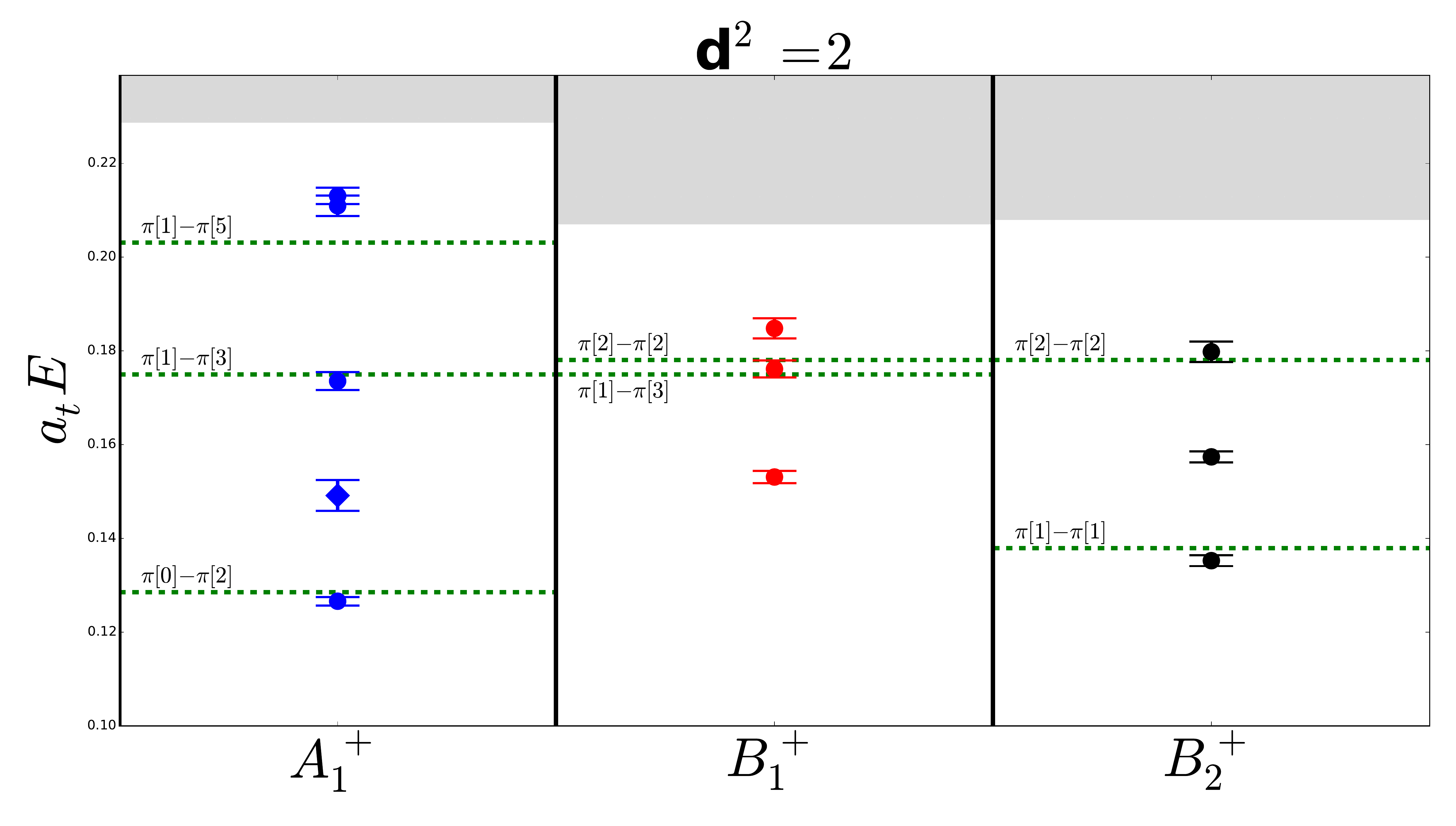}\\
  \includegraphics[width=2.9in]{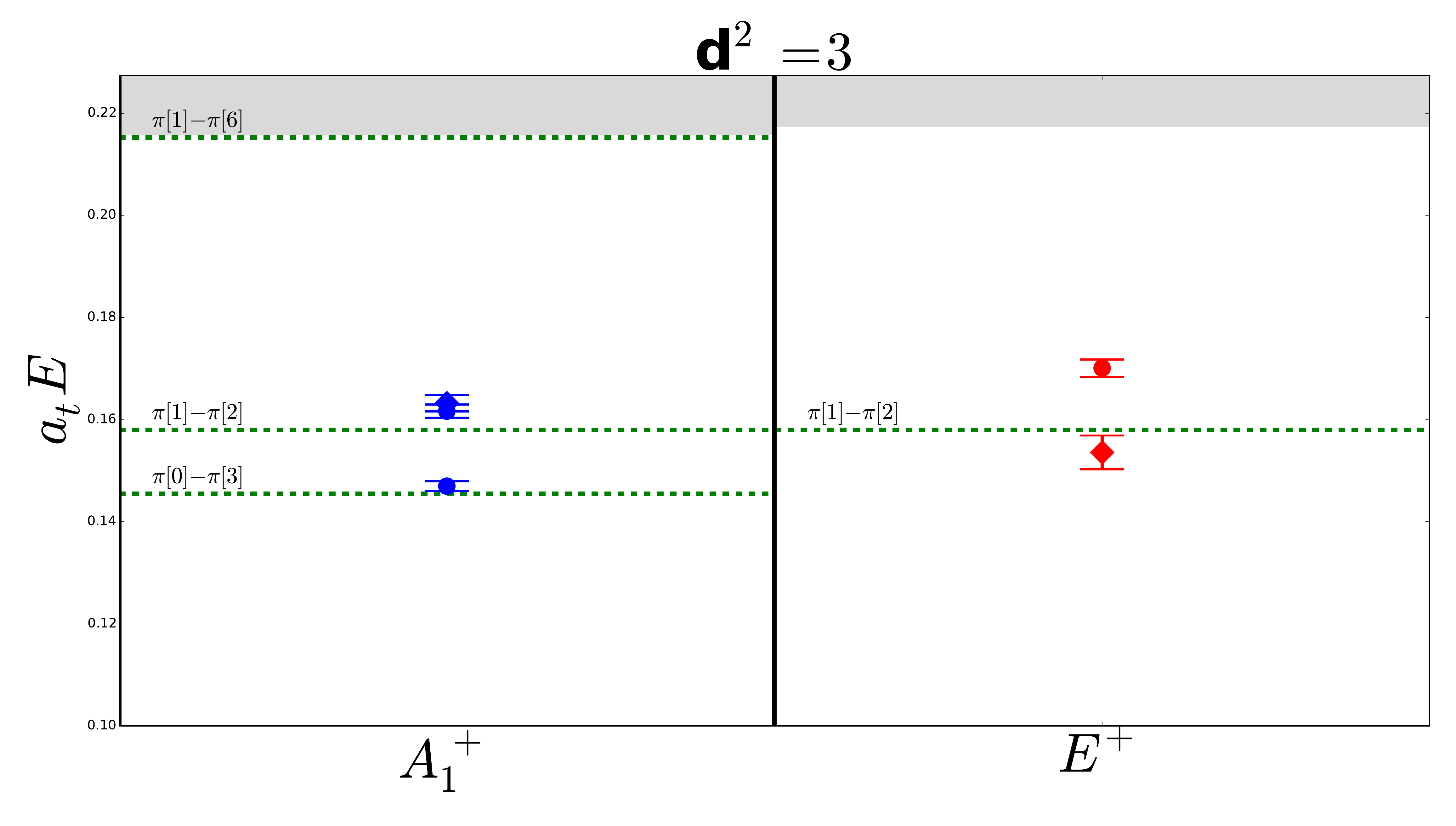}
  \includegraphics[width=2.9in]{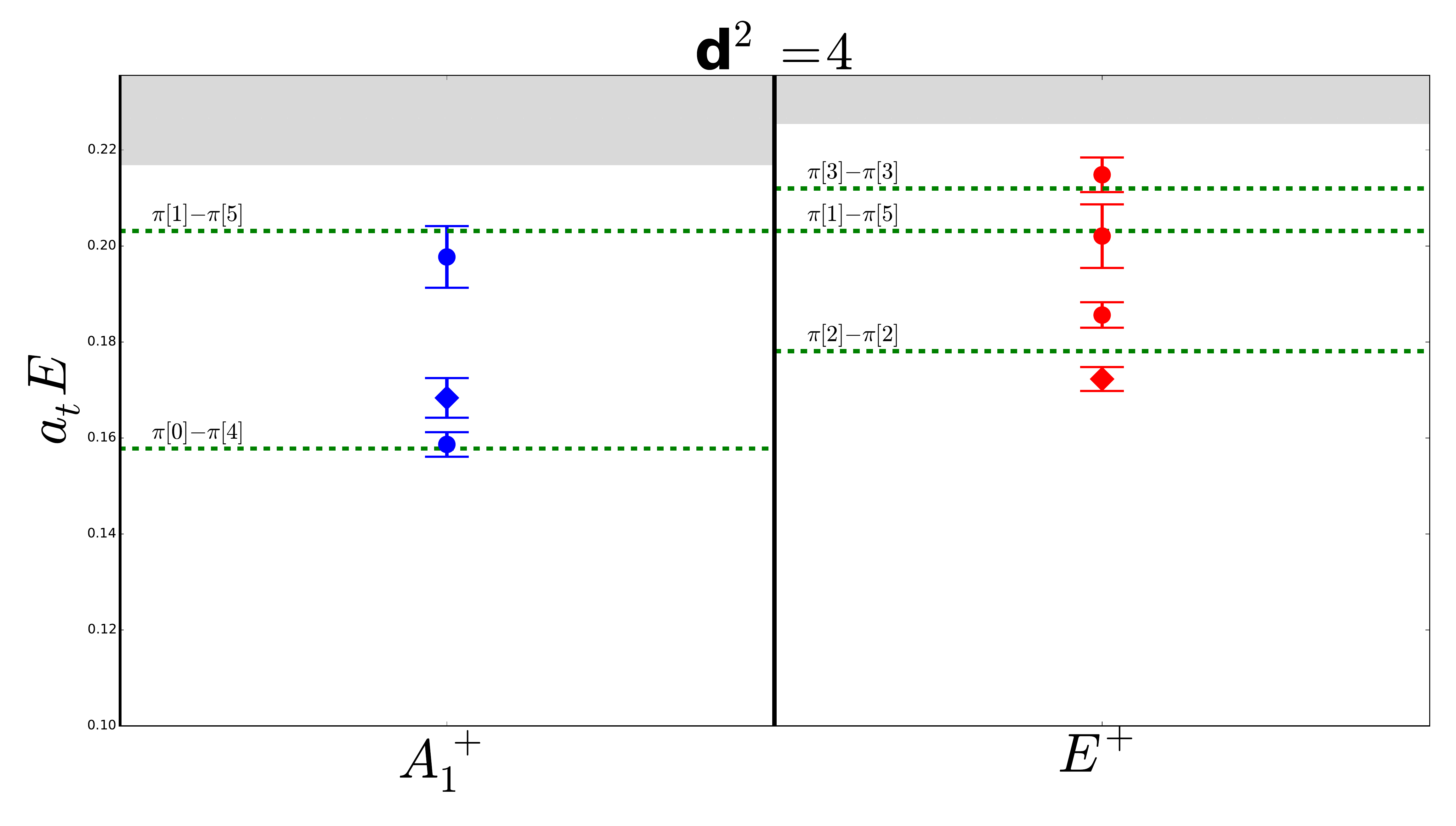}
  \end{center}
  \caption[Energies]{
    Energies $a_tE$ of $\pi\pi$ states for various $\dvec^2$. Dashed lines indicate the
    non-interacting energies of allowed $\pi\pi$ states. The shaded
    region indicates the inelastic thresholds. Diamond markers indicate
   levels with large overlaps with the $q\bar{q}$ operator corresponding to
  the $\rho$ meson.}
  \label{fig:pipispectra}
\end{figure}

\section{Finite-volume $\pi\pi\ I=1$ energies}
At rest, the $\rho$ meson appears in the $T_{1u}^+$ channel, but
for nonzero total momenta, we use results in Ref.~\cite{ExtendedHadrons}
to determine which little groups contain the $\rho$. We find that the
$\rho$ will appear in the irreps $A_1^+$ and $E^+$ of $C_{4v}$ for on-axis
total momenta, in the $A_1^+$,
$B_1^+$ and $B_2^+$ irreps of $C_{2v}$ for planar-diagonal momenta, and
$A_1^+$ and $E^+$ irreps of $C_{3v}$ for cubic-diagonal momenta. The
spectrum of energies from each of these channels can be used to
compute the $I=1$ $\pi\pi$ $P$-wave scattering phase shift, and hence,
determine the mass and width of the $\rho$ resonance.

In determining the $\pi\pi$ scattering phase shifts, only energy levels
below the inelastic thresholds can be used. In each of the above channels, we
include enough two-pion operators of different individual momenta to
get a good signal for all states below such thresholds.  Our operator
construction is described in detail in Ref.~\cite{ExtendedHadrons},
and our operator selection procedure and correlator matrix analysis
is presented in Ref.~\cite{lat2014spectrum}.  Fig.~\ref{fig:pipispectra} 
shows the energies obtained for the interacting $\pi\pi$ levels, compared 
to the energies of allowed $\pi\pi$ states in the absence of meson-meson 
interactions.  These results are obtained using a $32^3\times 256$ 
anisotropic lattice with quark masses tuned to yield a pion mass around 
240~MeV.  All needed Wick contractions were efficiently evaluated using
the stochastic LapH method\cite{StochasticLaph}.

\begin{figure}[t]
  \begin{center}
  \includegraphics[width=4.0in]{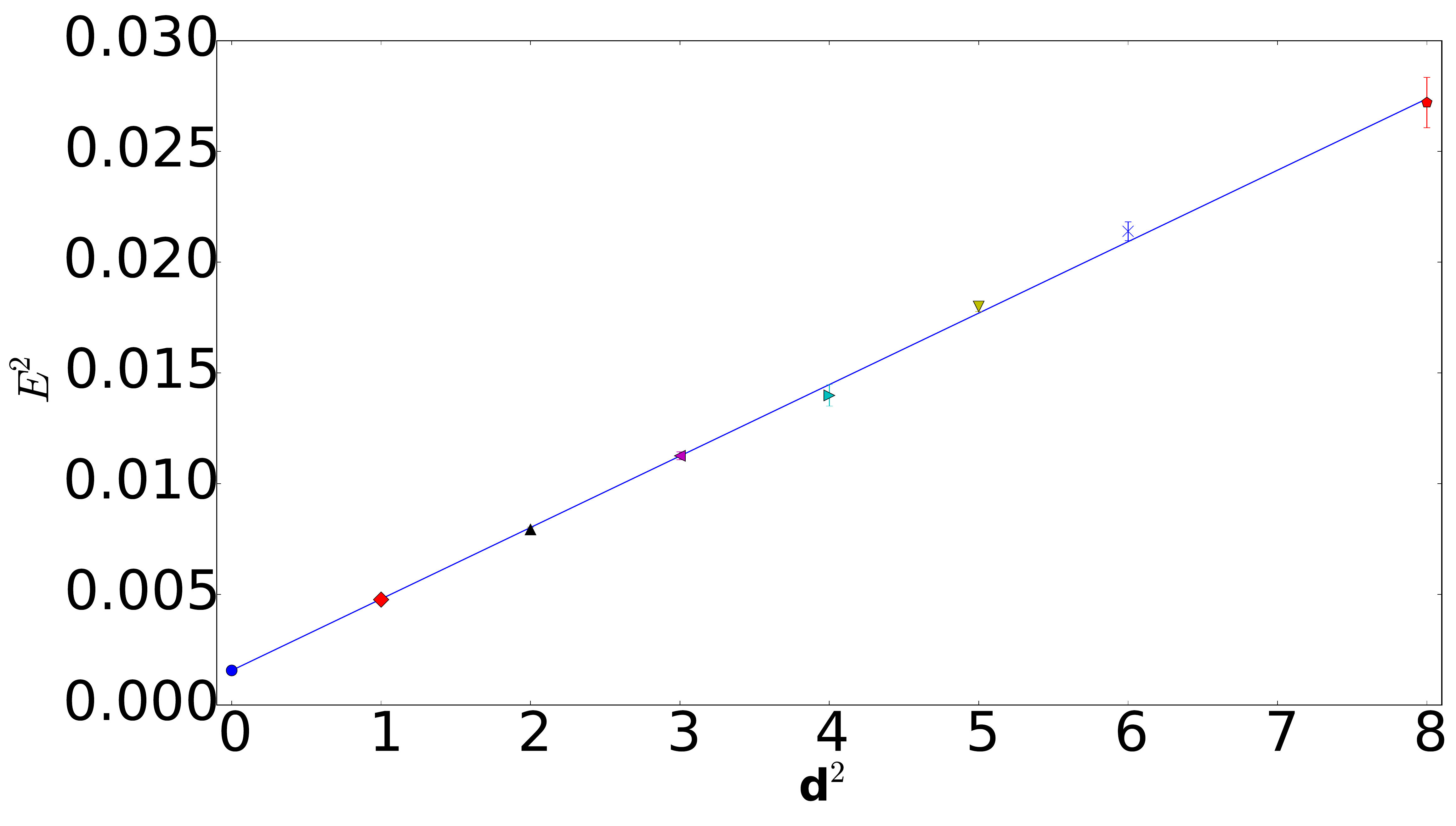}
  \end{center}
  \caption[Pion dispersion relation]{Single pion energies squared 
  $a_t^2E^2$ against $\dvec^2$ for our $32^2\times 256$ lattice.  Fitting to a 
  straight line yields the slope which equals $(\pi/(16\xi))^2$, where $\xi = a_s/a_t$.
  \label{fig:aspectratio}}
\end{figure}

\subsection{$P$-wave scattering phase shifts}
\label{sec:phaseresults}

To compute the scattering phase shifts using the energies for nonzero
total momenta, transformation to the center-of-mass frame is required. Since we are
using an anisotropic lattice, energies are measured in terms of the temporal 
spacing $a_t$, while the momenta are given in terms of the larger spatial spacing 
$a_s$. This means changing frames requires a precise knowledge of the 
renormalized anisotropy $\xi=a_s/a_t$.

We determine the anisotropy using the dispersion relation of the pion.
The energy $E$ of a free particle of mass $m$ and momentum $\Pvec=(2\pi/L)\dvec$ 
are related by
\beq
  \label{eq:dispersion}
  (a_t E)^2 = (a_t m)^2 + \frac{1}{\xi^2}\left(\frac{2\pi a_s}{L}\right)^2 
\dvec^2.
\eeq
By evaluating the energies of a particle with different momenta, $\xi$ can
be determined.  The energies for a single pion for various momenta are shown in
Fig.~\ref{fig:aspectratio}.  The parameter $\xi$ was fit using a
standard least squares fit for each bootstrap resampling.

\begin{figure}[t]
  \begin{center}
  \includegraphics[width=\textwidth]{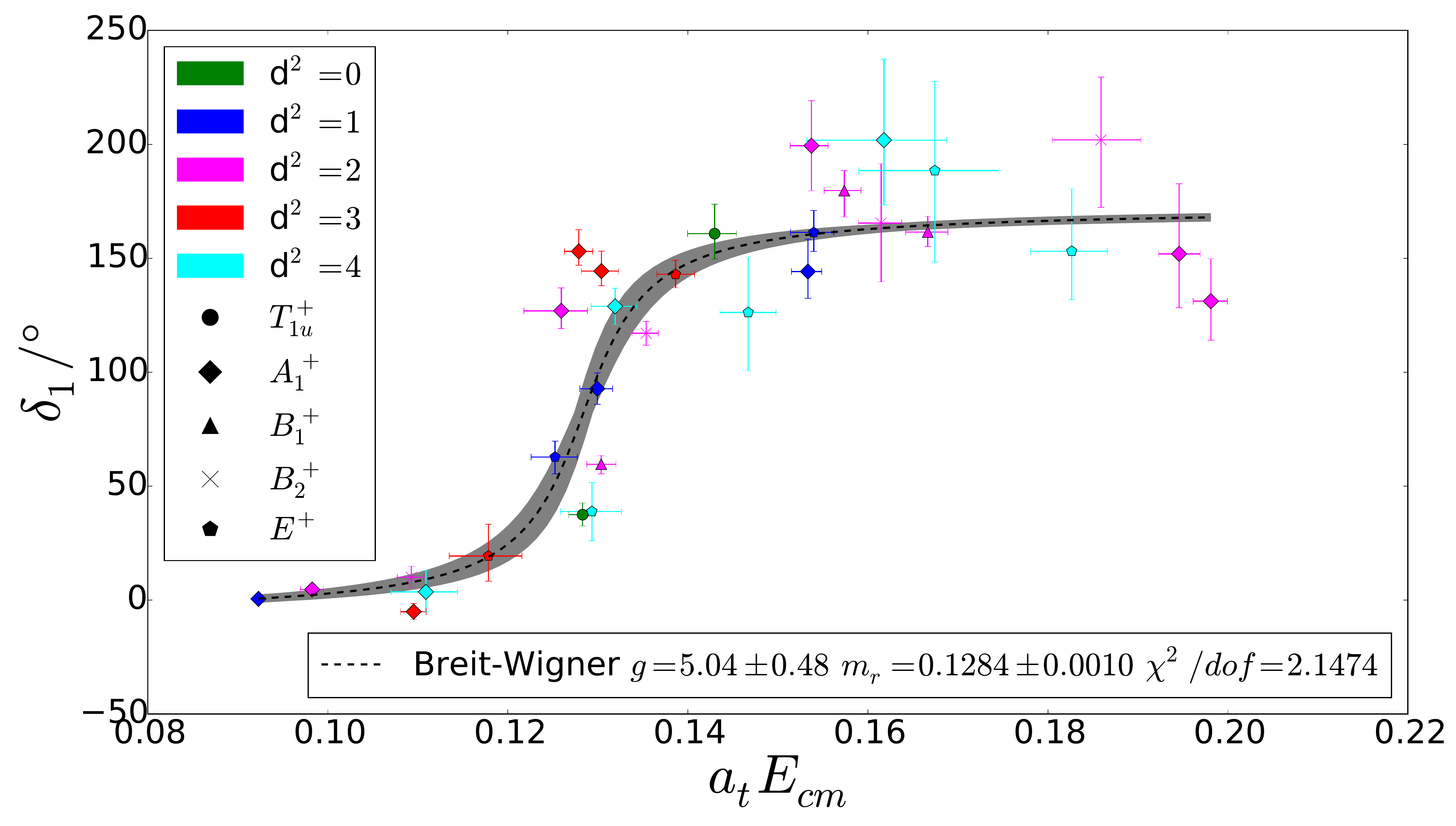}
  \end{center}
  \caption[$I=1$, $\pi\pi$ scattering phase shifts]{
    $P$-wave phase shift $\delta_1$ again center-of-mass energy $a_t E_{\rm cm}$. 
    Symbol color indicates the $\dvec^2$ of the level used, while the symbol shape indicates
    the irrep. Points which overlap $0^\circ$ or $180^\circ$ within
    error are shifted by $\pm 180$ to ensure continuity of the phase shift. The
    dashed line indicates the best fit to a Breit-Wigner with the gray
    band indicating the bootstrap errors in the fit function.  These results
    are still preliminary.
  \label{fig:phaseshifts}}
\end{figure}

The energies shown in Fig.~\ref{fig:pipispectra}, as well as the lowest 
three energies for zero momentum in the $T_{1u}^+$ channel obtained in 
Ref.~\cite{lat2014spectrum}, were used to compute the $\delta_1$ phase shift 
using the expressions given in Table~\ref{tab:cotdelta}.  Calculating the phase 
shift requires not only the energy $E$ of a particular state, but also the mass 
of the pion $m_\pi$ at rest and the renormalized anisotropy $\xi$ to determine
$\Ecm$, and hence, $\qcm$ and $u$. 
The formulas in Table~\ref{tab:cotdelta} yield $\cot{\delta_1}$, which
means that care with respect to quadrant must be taken when
determining $\delta_1$ for measurements on different bootstraps.

Our preliminary results for the $I=1\ \pi\pi$ $P$-wave scattering phase
shift are shown in Fig.~\ref{fig:phaseshifts} against the center-of-mass
energy $a_t \Ecm$.  The mass $m_r$ and width $\Gamma$ of the $\rho$ resonance 
is obtained by fitting the phase shift to a Breit-Wigner form:
\beq
  \label{eq:BW}
  \tan(\delta_1) = \frac{\Gamma/2}{ m_r - E } + A,
\eeq
where $A$ parametrizes a slowly-varying background.
The width $\Gamma$ is sensitive to the allowed phase space for its
decay products, which depends on the pion mass.  Since our pion mass is 240~MeV,
we cannot expect our width determination to agree with experiment.  However,
the effects of phase space can be reduced by writing the width
in terms of a $\rho\pi\pi$ coupling $g$:
\beq
  \Gamma = \frac{g^2}{48\pi m_r^2}\ (m_r^2 - 4m_\pi^2)^{3/2}.
\eeq
The coupling $g$ is expected to be fairly insensitive to the quark mass.
Our best-fit values for $m_r$ and $\Gamma$, with errors determined by
bootstrap resampling, are
\beq
  \label{eq:bwfitvalues}
  a_t m_r = 0.1284 \pm 0.0010 \quad \mbox{and} \quad g = 5.04\pm 0.48.
\eeq
The location of the resonance is consistent with the value obtained
from the spectrum of states in a finite box in the $T_{1u}^+$ channel
presented in Ref.~\cite{lat2014spectrum}, which yielded a mass
$0.1284 \pm 0.0014$. The value of $g$ is slightly low but consistent
with its experimental value near 6.  Keep in mind that these results
are not yet finalized. 

\section{Conclusion}
Our progress in calculating the $I=1$ $\pi\pi$ $P$-wave scattering phase shifts
on a large $32^3\times 256$ lattice for a light pion mass near 240~MeV was
described in this talk. The stochastic LapH method was used to evaluate
all needed Wick contractions.

This work was supported by the U.S.~National Science Foundation under awards 
PHY-1306805 and PHY-1318220, and through the NSF TeraGrid/XSEDE resources provided by 
TACC and NICS under grant number TG-MCA07S017. B.~H.\ is supported by Science 
Foundation Ireland under Grant No. 11/RFP/PHY3218.

\end{document}